\documentclass{aa}  

\usepackage{graphicx}
\usepackage{txfonts}
\usepackage{tabularx}
\usepackage{multirow}
\usepackage{booktabs}
\usepackage{float}
\usepackage{xcolor}
\usepackage{comment}
\usepackage{siunitx}
\bibpunct{(}{)}{;}{a}{}{,} 
\usepackage[breaklinks, colorlinks, citecolor=blue, linkcolor=blue]{hyperref}
\makeatletter
\renewcommand*\aa@pageof{, page \thepage{} of \pageref*{LastPage}}
\makeatother

\newcommand{\PlatoSim}{\texttt{PlatoSim} \,}

\usepackage[normalem]{ulem}

\begin{document}

\title{Detecting and sizing the Earth with PLATO: A feasibility study based on solar data}

\author{A. F. Krenn\inst{1} $^{\href{https://orcid.org/0000-0003-3615-4725}{\includegraphics[scale=0.5]{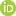}}}$\and
M. Lendl\inst{2} $^{\href{https://orcid.org/0000-0001-9699-1459}{\includegraphics[scale=0.5]{figures/orcid.jpg}}}$\and 
S. Sulis \inst{3} $^{\href{https://orcid.org/0000-0001-8783-526X}{\includegraphics[scale=0.5]{figures/orcid.jpg}}}$\and 
M. Deleuil\inst{3} $^{\href{https://orcid.org/0000-0001-6036-0225}{\includegraphics[scale=0.5]{figures/orcid.jpg}}}$\and 
S. J. Hofmeister\inst{4, 5} $^{\href{https://orcid.org/0000-0001-7662-1960}{\includegraphics[scale=0.5]{figures/orcid.jpg}}}$\and 
N. Jannsen\inst{6} $^{\href{https://orcid.org/0000-0003-4670-9616}{\includegraphics[scale=0.5]{figures/orcid.jpg}}}$\and
L. Fossati\inst{1} $^{\href{https://orcid.org/0000-0003-4426-9530}{\includegraphics[scale=0.5]{figures/orcid.jpg}}}$\and
J.~De~Ridder\inst{6} $^{\href{https://orcid.org/0000-0001-6726-2863}{\includegraphics[scale=0.5]{figures/orcid.jpg}}}$\and
D. Seynaeve \inst{6} $^{\href{https://orcid.org/0000-0002-0731-8893}{\includegraphics[scale=0.5]{figures/orcid.jpg}}}$\and
R. Jarolim \inst{7,8} $^{\href{https://orcid.org/0000-0002-9309-2981}{\includegraphics[scale=0.5]{figures/orcid.jpg}}}$\and 
A. M. Veronig \inst{7,9} $^{\href{https://orcid.org/0000-0003-2073-002X}{\includegraphics[scale=0.5]{figures/orcid.jpg}}}$
}

\authorrunning{A. F. Krenn et al.}

\institute{
\label{inst:1} Space Research Institute, Austrian Academy of Sciences, Schmiedlstrasse 6, A-8042 Graz, Austria \\
\email{andreas.krenn@oeaw.ac.at} 
\and
\label{inst:2} Observatoire Astronomique de l'Université de Genève, Chemin Pegasi 51, 1290 Versoix, Switzerland \and
\label{inst:3} Aix Marseille Univ, CNRS, CNES, LAM, 38 rue Frédéric Joliot-Curie, 13388 Marseille, France \and
\label{inst:4} 	Columbia Astrophysics Laboratory, Columbia University, 550 West 120th Street, New York, New York 10027, USA \and
\label{inst:5} 	Leibniz Institute for Astrophysics Potsdam, An der Sternwarte 16, 14478 Potsdam, Germany \and
\label{inst:6} 	Institute of Astronomy, KU Leuven, Celestijnenlaan 200D bus 2401, 3001 Leuven, Belgium \and
\label{inst:7} 	Institute of Physics, University of Graz, Universitätsplatz 5, 8010 Graz, Austria \and
\label{inst:8} High Altitude Observatory, National Center for Atmospheric Research, 3080 Center Green Dr, Boulder, CO 80301 \and
\label{inst:9} 	Kanzelh\"ohe Observatory for Solar and Environmental Research, University of Graz, Kanzelh\"ohe 19, 9521 Treffen, Austria
}

\date{Received 10.05.2024; accepted 10.10.2024}

 
  \abstract
  {The PLAnetary Transits and Oscillations of stars (PLATO) mission will observe the same area of the sky continuously for at least two years in an effort to detect transit signals of an Earth-like planet orbiting a solar-like star.}
   {We aim to study how short-term solar-like variability caused by oscillations and granulation would affect PLATO's ability to detect and size Earth if PLATO were to observe the Solar System itself. We also compare different approaches to mitigate noise caused by short-term solar-like variability and perform realistic transit fitting of transit signals in PLATO-like light curves.}
   {We injected Earth-like transit signals onto real solar data taken by the Helioseismic and Magnetic Imager (HMI) instrument on board the Solar Dynamics Observatory (SDO). We isolated short-term stellar variability in the HMI observations by removing any variability with characteristic timescales longer than five hours using a smooth Savitzky-Golay filter. We then added a noise model for a variety of different stellar magnitudes computed by \PlatoSim assuming an observation by all 24 normal cameras. We first compared four different commonly used treatments of correlated noise in the time domain by employing them in a transit fitting scheme. We then tried to recover pairs of transit signals using an algorithm similar to the transit least squares algorithm. Finally, we performed transit fits using realistic priors on planetary and stellar parameters and assessed how accurately the pair of two injected transits was recovered.}
   {We find that short-term solar-like variability affects the correct retrieval of Earth-like transit signals in PLATO data. Variability models accounting for variations with typical timescales at the order of one hour are sufficient to mitigate these effects. We find that when the limb-darkening coefficients of the host star are properly constrained, the impact parameter does not negatively affect the detectability of a transit signal or the retrieved transit parameters, except for high values ($b > 0.8$). For bright targets (8.5 - 10.5 mag), the transit signal of an Earth analogue can reliably be detected in PLATO data. For faint targets a detection is still likely, though the results of transit search algorithms have to be verified by transit-fitting algorithms to avoid false positive detections being flagged. For bright targets (V-mag $\leq$ 9.5), the radius of an Earth-like planet orbiting a solar-like star can be correctly determined at a precision of 3\% or less, assuming that at least two transit events are observed and the characteristics of the host star are well understood.}
   {}


   \maketitle
%


\section{Introduction}
While both planets with densities similar to Earth and planets orbiting solar-like stars have been previously detected, no Earth-like planet ($R_p < 1.2 \, R_{\oplus}$) orbiting in the habitable zone of a solar-like star (orbital periods between 200 and 500 days) has been found so far. Such a planet would be a prime target to search for signs of life, such as the presence of molecular oxygen in the atmosphere or the vegetation red edge \citep{Seager2005,Schwieterman2018,Meadows2018}. The detection of an Earth analogue is especially challenging due to the small signals planets similar to the Earth cause in well-established exoplanet detection methods such as the transit method ($\sim 80$ ppm) or the radial velocity method ($\sim 0.1$ m/s). The Characterising Exoplanet Satellite \citep[CHEOPS;][]{Benz2021} has proven photometric observations are capable of detecting planetary signals as small as $\sim 20$ ppm \citep{Brandeker2022,Demory2023,Krenn2023}. High-precision photometric instruments are in fact no longer limited by photon or instrumental noise, but much rather by the effects of stellar variability. Several stellar surface phenomena such as spots, plages, flares, convection, and oscillations affect the effectiveness of transit detection and fitting algorithms. Based on observations of the Sun and observations of solar-like stars by the \textit{Kepler} space telescope \citep{Borucki2010,Howell2014}, \citet{Sulis2020} for example have shown that short-timescale stellar variability associated with granulation noise can reach amplitudes of several hundred parts per million in photometric observations of solar-like stars. They also point out the importance of correctly accounting for the effects of this variability when trying to retrieve precise transit parameters. 

The PLAnetary Transits and Oscillations of stars (PLATO) mission \citep{Rauer2014,Rauer2024} is a photometric space mission designed specifically to search for Earth-like planets orbiting in the habitable zone of bright (V-mag < 11) and nearby solar-like stars. Planned to launch at the end of 2026, it will observe the same part of the sky for at least two years in an effort to detect and characterise exoplanets down to the size of Earth. It is equipped with a total of 26 individual wide field telescope units, called cameras, mounted on a single optical bench: 24 normal cameras, which are grouped in four groups of six cameras each, and two fast cameras. Each group of normal cameras will provide a light curve at 25s cadence and will use a broad bandpass of $500$ to $1000$ nm. The fast cameras, one using a green-red and one using a red-infrared filter, will produce light curves at 2.5s cadence. The field of view of the individual groups overlaps in the centre of the field of view of the whole instrument. Targets in the centre of the field of view of the instrument will be observed by all 24 cameras simultaneously, while targets at the edge of the field of view will only be observed by a single group (i.e. six cameras). Details on the observing strategy and technical specifications of the instrument can be found in \citet{Pertenais2021} and \citet{ Rauer2024}. 

The signal-to-noise ratio of an individual transit event is strongly dependent on the magnitude of the observed target and therefore the photon noise. Additionally, stellar variability can further distort the shape of the transit feature \citep{Czsesla2009} or even mimic a transit-like signal. Depending on the observing strategy, PLATO might only observe as few as two distinct transit events of every Earth-like planet it detects. Therefore, it is not possible to rely on a statistical cancellation of variability effects by adding up several transit events as it is often done when analysing datasets of short period planets. Rather, baseline models must be used to mitigate these effects in the individual transit events. 

In this work we aim to study how short-term stellar variability caused by oscillations and granulation would affect PLATO's ability to detect and size Earth, if Earth itself would be observed by PLATO from a distant vantage point. Previous observations by the \textit{Kepler} Space Telescope have shown that the Sun is photometrically quiet compared to other stars with similar properties \citep{Jenkins2002,Koch2010,Gilliland2011,Howell2016}. This limits the use of the Sun as a laboratory for globally quantifying the impact of stellar activity on light curves. However, it is by far the best-studied star and the only star for which high-resolution images of the stellar surface are available. These can be used to create disk-resolved artificial transit simulations that fully and realistically account for stellar surface features. These in turn can be used to investigate the best modelling approaches to analyse disk-integrated photometric light curves. While the Sun may not be a perfect proxy for all solar-like stars, insights gained from solar observations can be used on solar-like stars showing comparable variability levels as the Sun.

To this end, we created artificial PLATO-like transit light curves of an Earth-like planet by using real solar data to reproduce stellar noise and the most up to date instrumental noise calculations for PLATO employed in the \PlatoSim simulator (see Sect. \ref{sec_lcs}). We compare four different widely used treatments of correlated noise when applied to these light curves in Sect. \ref{sec_detrending}. In Sect. \ref{sec_detection} we also look at the performance of a transit search algorithm in detecting Earth-like transits in simulated PLATO data. Finally, in Sect. \ref{sec_fitting} we explain how we quantify the accuracy of retrieved transit parameters by performing transit fits of simulated PLATO data with realistic priors on stellar parameters.


\section{Simulated PLATO light curves of the Sun}
\label{sec_lcs}

To generate realistic PLATO-like transit light curves we followed a two steps approach. First we generated a realistic astrophysical signal. Then we injected it in the most up to date PLATO instrument simulator.

\subsection{Earth-like transit signal in solar observations}
\label{sec_solartransit}

For the astrophysical signal, we closely followed the approach presented in \citet{Sulis2020}, which uses solar observations taken by the Helioseismic Imager \citep[HMI;][]{Scherrer2012} instrument on board the Solar Dynamics Observatory \citep[SDO;][]{Pesnell2012}. HMI is a polarimetric Fabry-Perot instrument designed to measure the doppler velocities and magnetic field strength in the photospheric Fe I 6175 A line. It provides high-resolution images of the solar surface at 45s cadence. For each image pixel, it fits a Gaussian to the measured spectral profile from which the Stokes Vectors, Doppler shifts, line widths, and line depths of Fe I, and the adjacent continuum intensity is extracted. By integrating HMI's measured continuum intensity over the entire solar surface, we obtain the total solar flux. The so-derived solar flux resembles well the solar total irradiance measured by the Variability of Solar Irradiance and Gravity Oscillations (VIRGO) instrument in its green and red channels at \SI{500}{nm} and \SI{862}{nm} \citep{Sulis2020}, and thus should also be a good approximation of the flux that would be seen by PLATO's broadband filter, spanning a range of \SIrange{500}{1000}{nm}.

\begin{figure}
    \centering
    \includegraphics[width=\hsize]{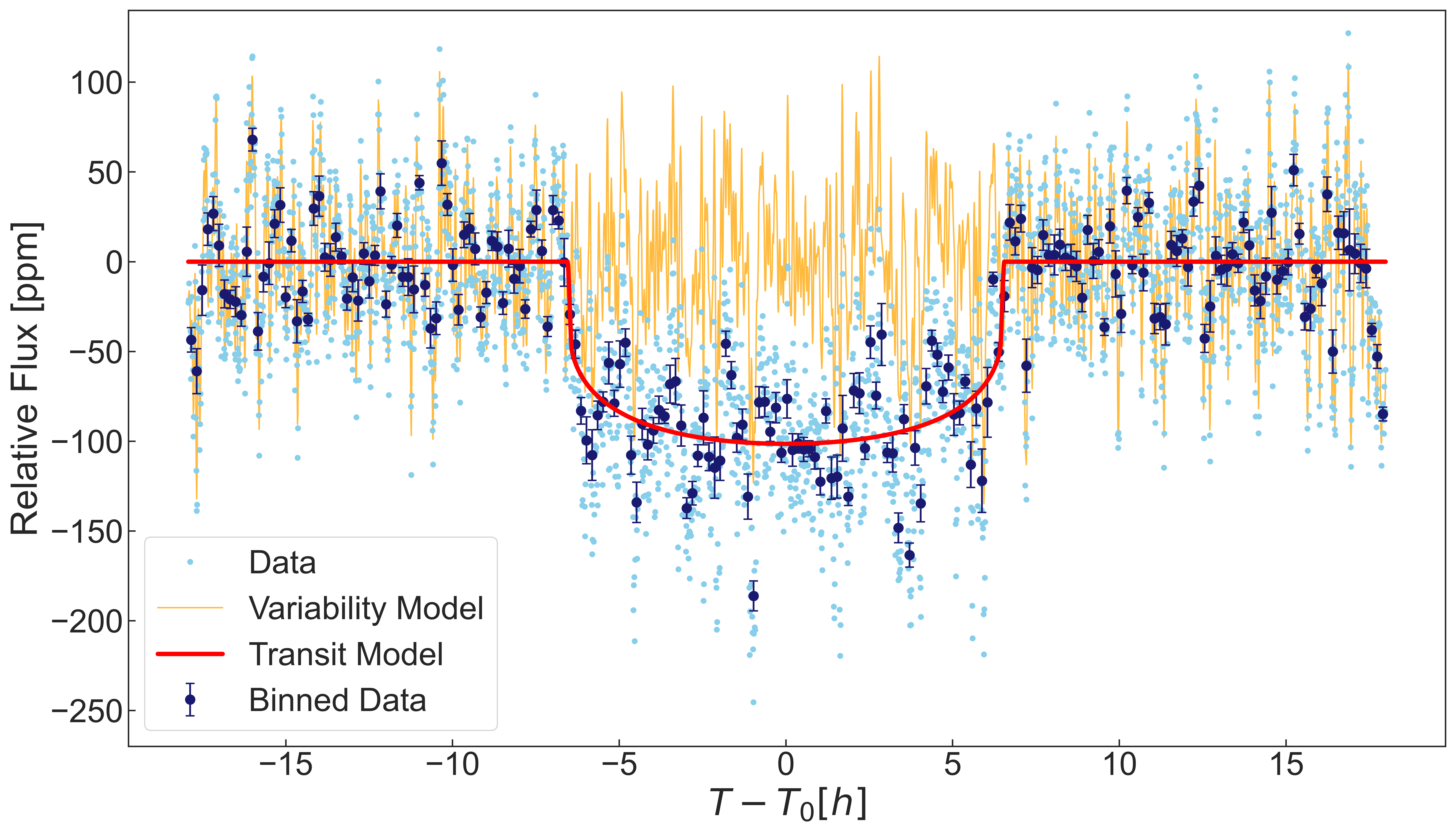}
    \includegraphics[width=\hsize]{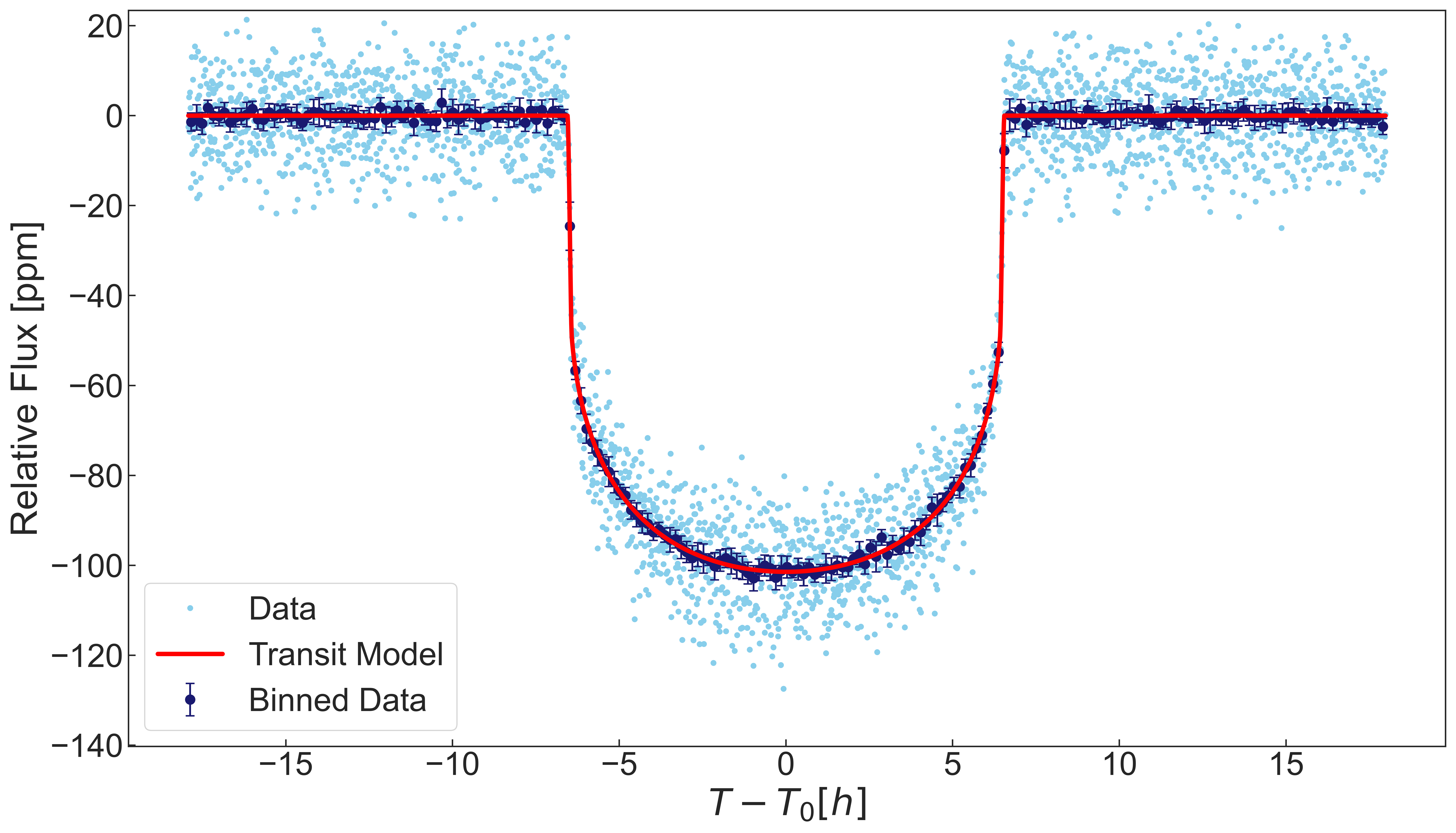}
    \caption{SDO HMI data (blue) of 17 October 2014 after correcting the data for the spacecraft motion and including the transit of an Earth-sized planet centred in the middle of the dataset. 10min-binned data (dark blue), injected transit shape (red), and assumed baseline (orange) when employing an SHO-GP to model stellar variability. Top panel: Raw data and fitted variability model. Bottom panel: The same data after correcting for stellar variability.}
    \label{fig:sho}
\end{figure}

\begin{figure}
    \centering
    \includegraphics[width=\hsize]{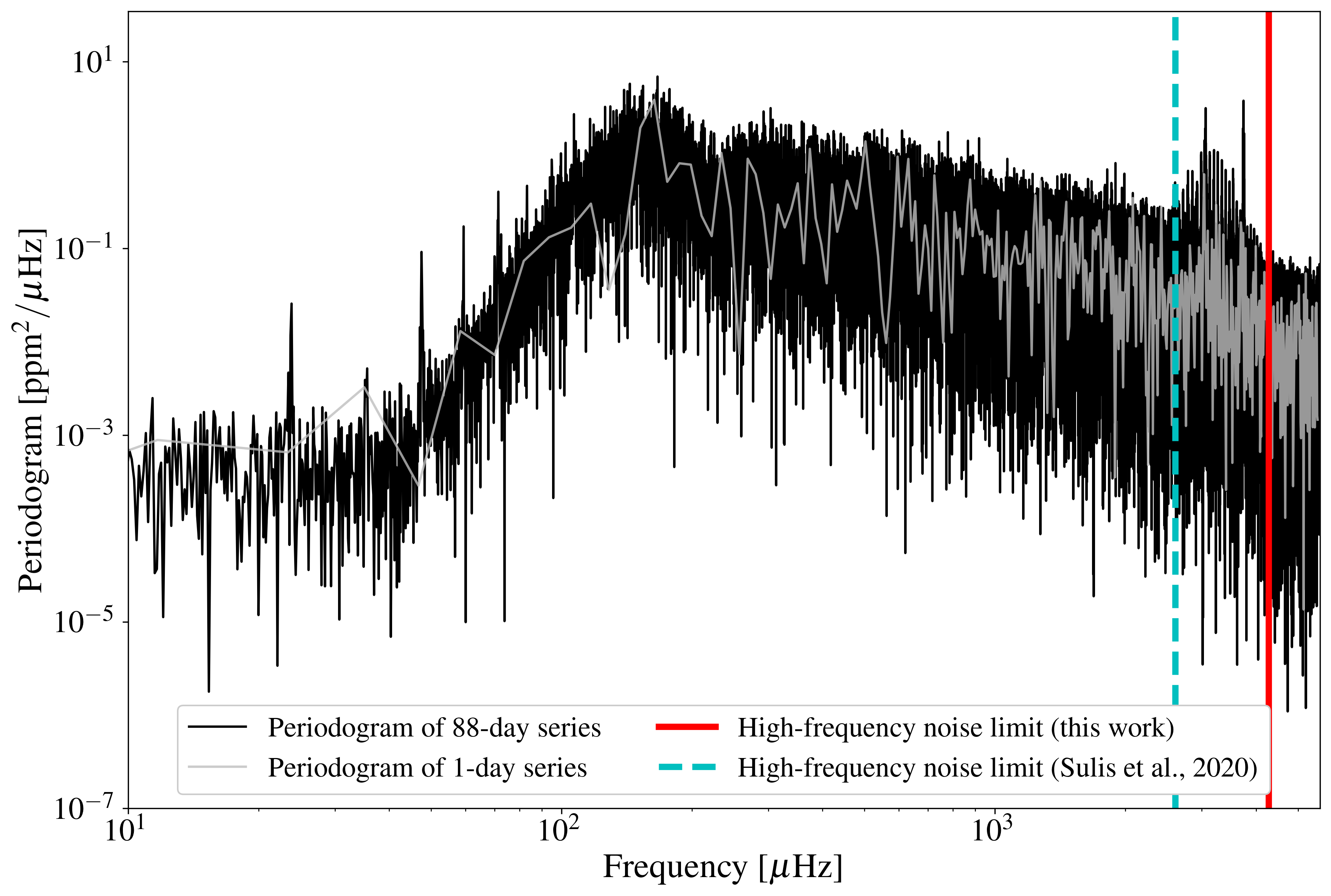}
    \caption{Periodogram of the 88-day HMI time series (black), with the periodogram of 1-day subseries shown for comparison (grey). The high-frequency noise of 14.8 ppm is estimated by taking the square root of the periodogram within the frequency range from 4275 $\mu$Hz (red line) to the Nyquist frequency. The dashed cyan line marks the 2604 $\mu$Hz frequency discussed in  footnote 1. The effect of the Savitzky-Golay filter is clearly shown in the periodogram at frequency $<$ 100 $\mu$Hz.}
    \label{fig:periodo}
\end{figure}

The artificial transit light curve of the exoplanet is created by superimposing a black sphere on the HMI images and moving it across the disk. In our study, we oversampled the exoplanet mask by a factor of 11, which is necessary to allow partially eclipsed pixels and therefore a high-precision of the occulting mask. We assumed a planet size $R_p = 1$ $R_{\oplus}$, an orbital period of $P = 365.25$ days and simulated transits for impact parameters $b = 0, 0.25, 0.5, 0.75$, and $0.9$. The relative velocity of SDO with respect to the Sun is not constant but varies at a period of $\sim$ one day due to the spacecrafts motion along its orbital trajectory in low-Earth-orbit. This results in a variation of the observed flux, because of a slight Doppler shift of the observed wavelength range of solar light. Analogous to \citet{Sulis2020}, we accounted for this effect by applying a smooth Savitzky-Golay filter to the raw (i.e. without a transit) light curves. The filter removes temporal trends with characteristic time scales longer than five hours. Since magnetic features such as plages and spots on the solar surface evolve at timescales much longer than five hours, flux variations caused by such features are removed from our dataset, meaning that the photometric effects of stellar activity are removed. Instead, the stellar variability that we observe in our data is dominated by short-term noise sources, most importantly surface granulation and oscillations (see Fig. \ref{fig:periodo}). During the solar activity cycle the number of magnetic features varies with the most features being observed at the maximum of the activity cycle. Flux features caused by oscillations and granulation are however mostly unaffected by the solar activity cycle. Because we remove flux features caused by magnetic features, the noise sources in our dataset are practically identical to the noise sources in the dataset used by \citet{Sulis2020}, although one study used data observed during the maximum and the other during the minimum of the solar activity cycle. We choose 88 consecutive days of solar observations during a solar maximum from 10th of October 2014 to 5th of January 2015. We centred one transit at the middle of each dataset associated with a calendar day (i.e. roughly at noon Universal Time of each day). This results in 88 individual transit events in the 88-day time series. As we simulated five different impact parameters, we ended up with five 88-day time series and 440 individual transit events (see Fig. \ref{fig:sho} as an example). 

To properly scale the typical uncertainties of individual HMI data points, which are not provided, we quantified the high-frequency noise in HMI observations. For that purpose, we computed the periodogram for the entire 88-day time series and evaluated the noise level at high frequencies, specifically between 4275 $\mu$Hz (indicated by the red vertical line in Fig. \ref{fig:periodo}) and the Nyquist frequency (0.01 Hz). The estimated noise amplitude in this range is 14.8 ppm\footnote{It is worth noting that our approach is more conservative than that of \citet{Sulis2020}, where the high-frequency noise was assessed using GP fits on 1-day time series, resulting in a noise amplitude of 25 ppm. In Fig. \ref{fig:periodo}, this higher noise level corresponds to the power in the frequency range between 2604 $\mu$Hz (marked by the cyan dashed line) and the Nyquist frequency. It includes the p-mode oscillations. However, the p-modes are unresolved in the 1-day time series analysis, as used both in the transit characterisation in \citet{Sulis2020} and in the present study.}, which was used as uncertainties for all HMI data points.

To quantify typical variations caused by short-term solar variability in the HMI data, we calculated the standard deviation of the whole 88-day times series without the artificial transits, which is 46 ppm. Furthermore, we attempted to model these variations by fitting a Simple-Harmonic-Oscillator Gaussian process \citep[SHO-GP;][]{Foreman-Mackey2017} model to the first ten days of the dataset without the artificial transit signal added. The mathematical description of the GP kernel can be found in Appendix \ref{sec_appendix_gp}. We retrieved the following GP hyperparameters:

\begin{multline*}
    \rm ln(S_0) = -26.79 \pm 0.04, \\
    \rm ln(Q) = -0.402 \pm 0.025, \\
    \rm ln(\omega_0) = 6.932 \pm 0.011. \\
\end{multline*} 

These parameters translate to a typical timescale of variations of $\sim$ 9 min and typical amplitudes of variations of $\sim$ 40 ppm. Fig. \ref{fig:sho} shows an example of a GP-baseline model with these hyperparameters applied to a dataset including an artificial transit.

\begin{table*}
    \caption{Overview of noise models used to detrend the simulated photometric transit observations.}
    \centering
    \begin{tabular}{cccc}
    \toprule
    \toprule
    Abbreviation & Name & Description & Source \\
    \midrule
    FB & Flat baseline &\parbox{9cm}{A flat (normalised flux always equal to 1) baseline was used} & -  \\
    \midrule
    P2 & Quadratic Polynomial & \parbox{9cm}{At each sampling step, a least squares minimisation determines the 2nd-order polynomial parameters setting the baseline} & -  \\
    \midrule
    SP & Spline & \parbox{9cm}{At each sampling step the residuals are described with a smooth cubic spline} & -  \\
    \midrule
    MGP & Matérn-3/2 GP & \parbox{9cm}{The baseline is determined by a Gaussian Process using a Matérn-3/2 Kernel} & \parbox{2.5cm}{\citet{Rasmussen2006}}  \\
    \midrule
    SGP & SHO GP & \parbox{9cm}{The baseline is determined by a Gaussian Process using a Simple Harmonic Oscillator Kernel} & \parbox{2.5cm}{\citet{Foreman-Mackey2017}}  \\
    \bottomrule
    \end{tabular}
    \label{table_detrending}
\end{table*}

\subsection{PLATO-like light curve simulations}

The final light curve products, which had the SDO spacecraft motion removed and the transits added, were then used as an input to the \PlatoSim simulator \citep{Jannsen2024}. This software package is specifically designed to produce PLATO-like noise. Using \PlatoSim's toolkit and \texttt{PLATOnium}, we selected one target in the PLATO input catalogue \citep{Montalto2021} that could be observed by all 24 cameras. We configured the simulator to switch on all basic instrumental effects, simulated light curves for 24 cameras, and averaged these to a single light curve. The simulation of an individual camera was reduced to a small $10 \times 10$ pixel window centred on the selected target. This allows to significantly decrease the calculation time, especially as we are interested in a single target only. Since PLATO will provide 25s cadence data, but HMI only provides 45s cadence data, the simulator was set to integrate measurements to 45s cadence. With this setup we then performed a series of simulations, varying both the V-band magnitude of the target star, selecting values of 8.5, 9.5, 10.5, 11.5, and 12.5, and the impact parameter of the transiting planet. We performed separate simulations for all combinations of magnitudes and impact parameters. Each time we simulated a full quarter (88 days) of observations. This resulted in a total of 25 88-day times series and 2200 individual transit events. 

The simulated light curves then contain solar flux features with characteristic timescales of less than five hours, the artificial exoplanet transit and realistic PLATO-like noise. The most prominent flux feature is a decline of flux, which is mostly due to ageing of the instrument. As an example of the final transit light curves, we have published a figure showing one day of the simulated transit light curves for all five magnitudes online \footnote{\url{https://zenodo.org/records/13938622}}.


We note that during the course of the actual mission, for bright stars for which imagettes will be available (V < 11), the photometry will be extracted on-ground using a PSF fitting algorithm, whereas, for fainter stars it will be done on board by aperture photometry \citep{Samadi2019,Marchiori2019}. As the implementation of the PSF-fitting method is still under development in the ground-segment pipeline, \PlatoSim currently provides only optimal aperture photometry. A detailed comparison of the performances of the two approaches for the specific case of  PLATO has not yet been published. Previous PSF fitting methods employed for space-based photometeric observations, such as PIPE for CHEOPS \citep{Brandeker2024}, have shown results to be of similar or better accuracy when using PSF fitting instead of aperture photometry. Therefore the use of the optimal aperture photometry can be seen as a conservative approach. However, in general PSF fitting only leads to better results when the aperture is strongly contaminated by background stars or stray light \citep{Brandeker2024}. Since we did not include any contaminating star in our simulations, we do not expect a significant difference between the two methods. The whole set of artificial transit light curves is publicly available online\footnote{\url{https://zenodo.org/records/13939257}}. 

\begin{figure}
    \centering
    \includegraphics[width=\hsize]{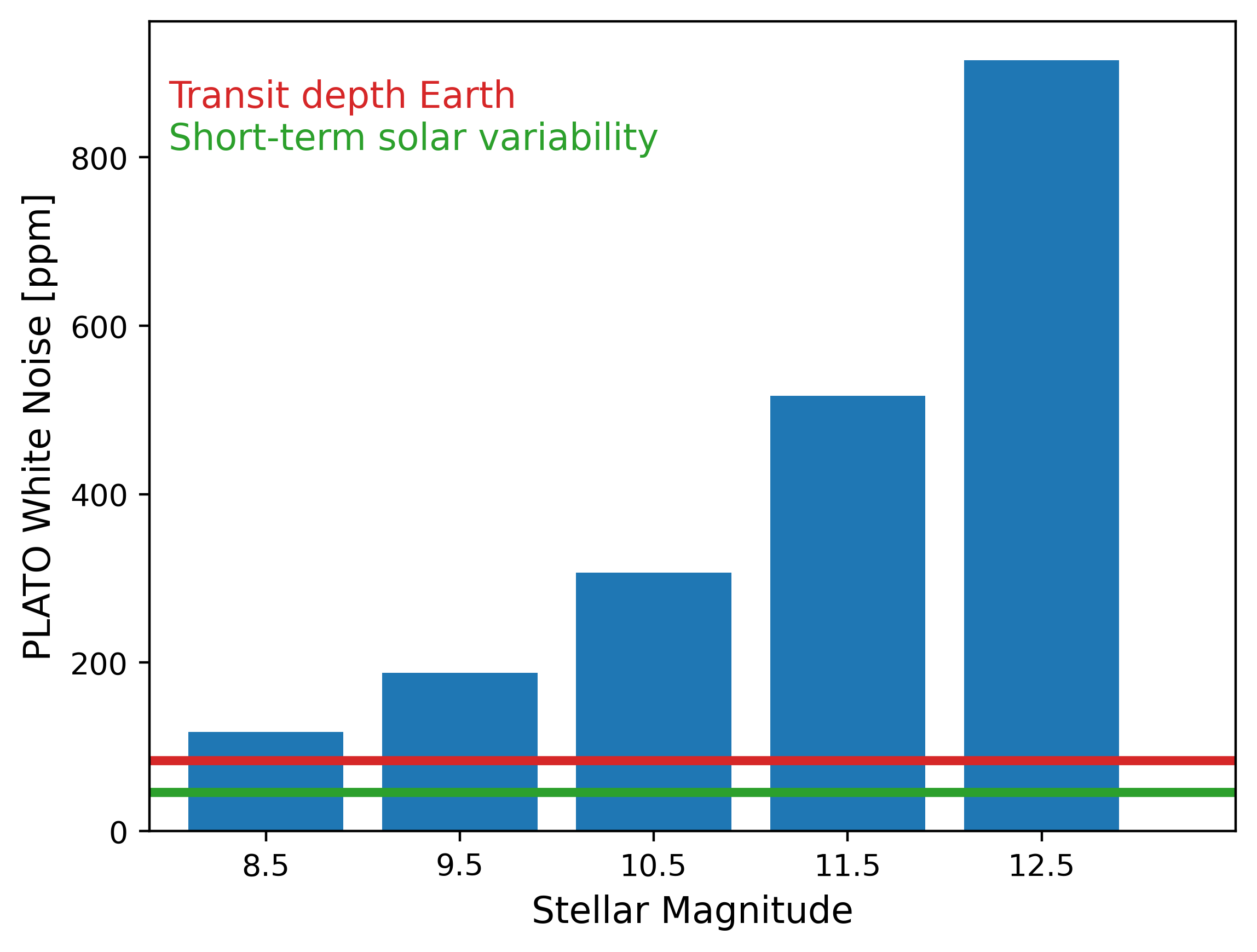}
    \caption{Overview of typical amplitudes of uncorrelated variations in the PLATO noise model (blue) for different stellar magnitudes. The red line represents the depth of the transit signal of an Earth-like planet orbiting a solar-like star (84 ppm). The green line represents typical amplitudes of variations caused by short-term solar variability (46 ppm).}
    \label{fig:noise}
\end{figure}

To assess the contributions of different noise sources to the PLATO noise model we fitted a combined model of a Gaussian Process using a Matern-3/2 kernel \citep{Rasmussen2006} and a white-noise term to the light curves produced, which assumed the planetary transit to occur on the equator of the star. Prior to fitting the model, we ensured for the data to only contain short-term PLATO-specific noise by removing the injected astrophysical signal (short-term stellar variability and planetary transits), performing outlier removal, and correcting for the characteristic PLATO slope with a second order polynomial in time. We find that for all stellar magnitudes, the remaining correlated noise is negligible with typical variations being two-orders of magnitude smaller than the uncorrelated variations. We therefore conclude that the short-term PLATO noise model provided by \PlatoSim is dominated by uncorrelated noise. The amplitudes of typical variations caused by this uncorrelated noise for different stellar magnitudes are shown in Fig. \ref{fig:noise}. We note that, while for bright targets this contribution to the overall noise budget is of similar order of magnitude as the planetary transit signal and the short-term stellar variability signal, for faint targets the white-noise contribution by the PLATO noise model is the dominating noise source.

\begin{table*}
    \caption{Model parameters used in the transit analysis when comparing different noise models.}
    \centering
    \begin{tabular}{lcccc}
    \toprule
    \toprule
    Parameter & Symbol & Prior & Correct value \tablefootmark{(a)} &  Units \\
    \midrule
    \textbf{Planet Parameters}&&&& \\
    \quad Orbital period & $P$ & $\mathcal{F}(365.25)$ & $365.25$ & days \\
    \noalign{\smallskip}
    \quad Transit time & $T_{0}$ & $\mathcal{U}(0.48,0.52)$ & 0.4995 - 0.5005 \tablefootmark{(b)} & days\\
    \quad Planet-to-star radius ratio & $R_{p}/R_\star$ & $\mathcal{U}(0.0,0.02)$ & $0.009168$  &- \\
    \quad Scaled semi-major axis & $(R_\star + R_p) / a$ & $\mathcal{U}(0.0,0.02)$ &  $0.004693$ &  - \\
    \noalign{\smallskip}
    \quad Cosine of inclination & $\rm cos(i)$ & $\mathcal{U}(0.0,0.006)$ &  varying &  - \\
    \quad White noise term & $\sigma_W$ & $\mathcal{U}(45,6738)$ &  varying &  ppm \\
    \noalign{\smallskip}
    \midrule
    \textbf{Stellar Parameters} &&&& \\
    \quad Stellar mass & $M_\star$ &$\mathcal{F}(1.0)$ & 1.0 &  $M_\odot$ \\
    \quad Stellar radius & $R_\star$ &$\mathcal{F}(1.0)$ & 1.0 &  $R_\odot$ \\
    \multirow{2}{*}{\quad limb-darkening coefficients} & $q_1$ &$\mathcal{N}(0.392,0.0085)$ & - &  - \\
    \noalign{\smallskip}
    & $q_2$ & $\mathcal{N}(0.322,0.0117)$ & - &  - \\
    \midrule
    \textbf{Matérn-3/2 GP} &&&& \\
    \quad Amplitude & $\rm ln(\sigma)$ &$\mathcal{U}(-20,-1)$ & - & - \\
    \quad Length scale parameter & $\rm ln(\rho)$ &$\mathcal{U}(-15,5)$ & - & - \\
    \midrule
    \textbf{SHO GP} &&&& \\
    \quad Scaled amplitude & $\mathrm{ln(S_0)}$ &$\mathcal{N}(-26.79,0.04)$ & - & - \\
    \quad Quality factor & $\mathrm{ln(Q)}$ &$\mathcal{N}(-0.402,0.025)$ & - & - \\
    \quad Angular Frequency & $\mathrm{ln}(\omega_0)$ &$\mathcal{N}(6.932,0.011)$ & - & - \\
    \bottomrule
    \end{tabular}
    \tablefoot{The Gaussian priors with mean $\mu$ and standard deviation $\sigma$ are displayed as $\mathcal{N}(\mu, \sigma)$. $\mathcal{U}(U_l,U_u)$ represents an uniform prior with $U_l$ and $U_u$ as lower and upper limits respectively. $\mathcal{F}$ describes a fixed value.\\
    \tablefoottext{a}{Value of the injected transit signal.}\\
    \tablefoottext{b}{The injected transit mid-time is always the exact middle of the dataset, which varies accross the different observations according to the exact observation times of the HMI images.}}
    \label{table:transit_parameters}
\end{table*}

\section{Comparing different noise models}
\label{sec_detrending}
Photometric observations of exoplanet transits are often affected by a variety of noise sources unrelated to the planetary transit itself. These can both be instrumental (i.e. impacts of high-energy particles, smearing noise on the CCD, etc.) and astrophysical, such as a contaminating star or stellar variability \citep{Christiansen2020}. To mitigate the effects of these noise sources a variety of different noise models are used in the analysis of transit observations. These methods vary substantially in the complexity of the employed algorithm, ranging from polynomial functions correlated to supplementary parameters (e.g. time) to Gaussian Processes \citep[GPs, see e.g.][]{Rasmussen2006,Gibson2012,Foreman-Mackey2017}. In this study, the time series describing the model of the correlated noise is called the baseline, because it represents the expected stellar flux if no transiting planet is present. To compare the performance of different noise models we performed a series of transit analyses employing five different treatments of time-correlated noise integrated in the transit-fitting software package \texttt{allesfitter}\footnote{ \url{https://github.com/MNGuenther/allesfitter}}  \citep{Guenter2021}: A flat baseline (FB), a second-order polynomial (P2), a spline function (SP), a Gaussian Process using a Matérn-3/2 Kernel \citep[MGP,][]{Rasmussen2006}, and a Gaussian Process using a Simple Harmonic Oscillator Kernel \citep[SGP,][]{Foreman-Mackey2017}. Detailed information on these methods can be found in Table \ref{table_detrending}. The mathematical description of the GP kernels can be found in Appendix \ref{sec_appendix_gp}. The parameters of the baseline models were fitted simultaneously with the transit parameters, except for the hyperparameters of the SGP model. Mimicking the approach of first determining the hyperparameters of the SHO-GP by analysing only out-of-transit data \citep{Krenn2023}, we imposed Gaussian priors on them according to the values retrieved when analysing the short-term solar variability in the HMI images (see Sec. \ref{sec_solartransit}). These priors, derived from high-resolution images of the solar surface, are strongly constraining. PLATO will utilise two years of consecutive observations of the host stars within its dedicated stellar oscillations programme to precisely determine the characteristics of the stellar oscillation and granulation patterns. These extensive analyses combined with knowledge gained from solar observations, will allow the PLATO team to put tight constrains on the characteristic amplitudes and time scales of short-term stellar variability in solar-like stars. Since the target star of the artificial transit light curves produced in this work is assumed to be identical to the Sun, such narrow priors are justified when mimicking the analysis of PLATO light curves.

\begin{figure*}
    \centering
    \includegraphics[width=\hsize]{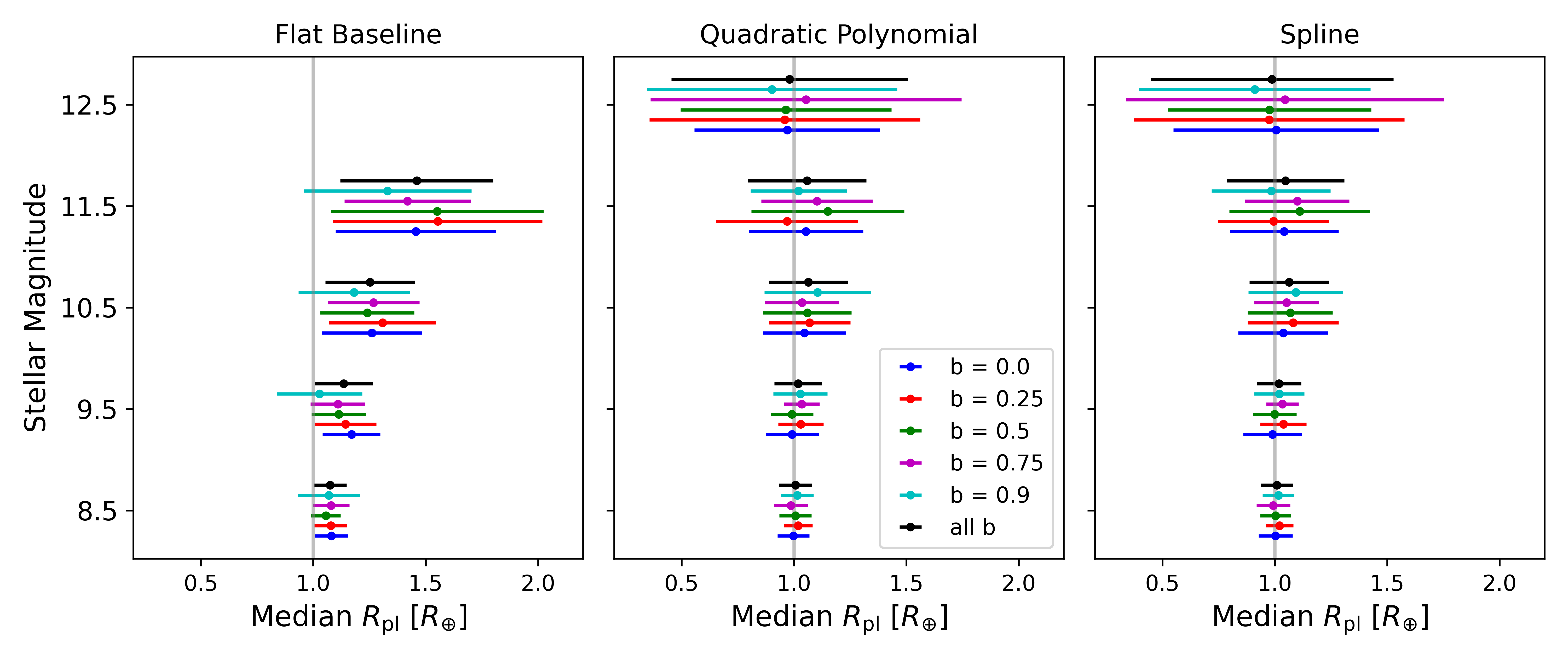}
    \includegraphics[width=0.66\hsize]{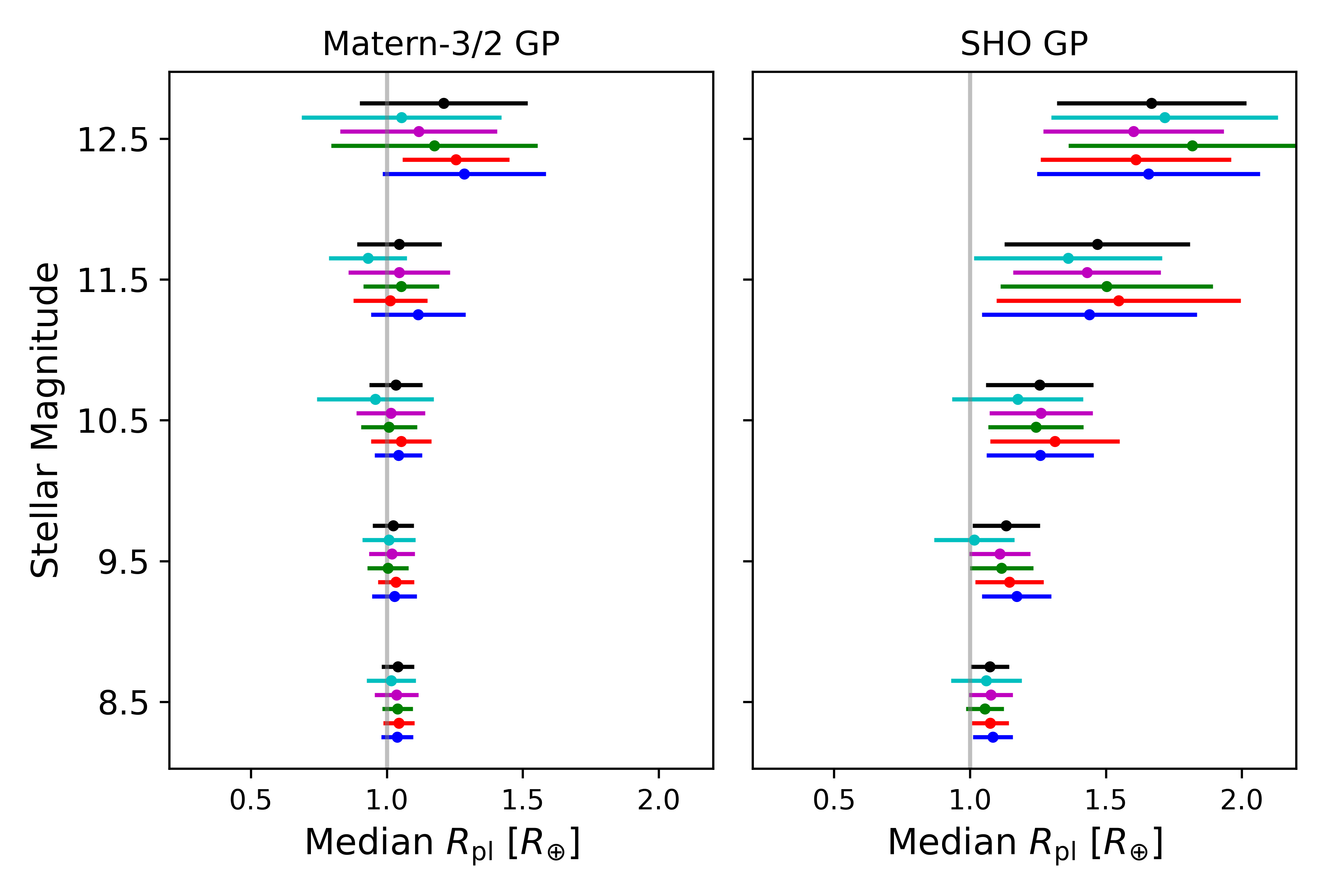}
    \caption{Overview of retrieved median planetary radii and their 1$\sigma$-confidence intervals as a function of the stellar magnitude and impact parameter (colour-coded) for different baseline models. Each point represents a distribution of 88 planetary radii, with each radius being retrieved from one of the 88 one-day light curves containing a single transit event. The grey vertical lines represent the injected signal. In the case of a flat baseline (FB) and stellar magnitude 12.5 the results could not be visualised because the majority of solutions are at the upper boundary of the parameter space.}
    \label{fig:baseline_overview}
\end{figure*}

Each of the available 88-day time series (see Sect. \ref{sec_lcs}) was split into 88 individual light curves containing each 24 hours of data and one single transit event centred in the middle of that dataset. Combining the 88 days, the five magnitudes, and the five impact parameters, this results in a total of 2200 light curves. For each of the five baseline models, a transit was fit to each individual light curve using the dynamic nested sampling algorithm \citep[e.g.][]{feroz2008,feroz2019} implemented in \texttt{allesfitter} via the \texttt{dynesty} Python package \citep{speagle}. The parameters of the transit model including the imposed priors are listed in Table \ref{table:transit_parameters}. To quantify the effects of different baseline models without biasing the results due to uncertainties in stellar parameters, we fixed both the stellar mass and radius to the solar values and used the stellar density to constrain the transit model. To account for the varying amplitudes of white noise for different stellar magnitudes, we also fitted for a separate white noise term. Additionally, we also used a quadratic limb-darkening law using coefficients following the parametrisation proposed by \citet{Kipping2013}. To determine the coefficients, we fitted them to the transit shapes retrieved when assuming a transit across the solar equator (i.e. an impact parameter of $b = 0$), while fixing all other transit parameters to the injected value. We then computed the mean and standard deviation of this set of 88 combinations of limb-darkening coefficients and imposed a Gaussian prior in the transit analysis accordingly. 

To compare the performance of the different noise models, we computed the weighted median and its 1$\sigma$-confidence interval for the distribution of retrieved planetary radii (see Table \ref{table_detrending_results} and Fig. \ref{fig:baseline_overview}). When a flat baseline is applied, we observe an overestimation of the correct planetary radius, with the overestimation being increasingly significant for fainter stars. In the case of the magnitude 12.5 target the retrieved radius ends up at the upper boundary of the allowed radius values for the majority of the fitting attempts, indicating that the algorithm would have overestimated the radius even more significantly if it would have been allowed to. Simple baseline models such as a polynomial or a spline are able to remove this systematic overestimation. However, with decreasing brightness, the distributions get much broader, indicating that individual transit events are still significantly under- or overestimated. The more complex baseline model MGP using a Matérn-3/2 Gaussian process does not show an overestimation bias. The retrieved radii distribution of MGP is narrower compared to P2 and SP, indicating that individual measurements are more accurate. Notably, the characteristic timescale of variations modelled by the MGP model is $\sim$1 h, implying that the Gaussian process is not accounting for the short-period part of the solar granulation and p-mode timescale spectrum at the order of 5 to 10 minutes \citep{Nesis2002,Sulis2020}. Similarly to the flat baseline, the SGP baseline model, which employs a Simple-Harmonic-Oscillator Gaussian process, also systematically overestimates the correct radius. The characteristic timescales of variations modelled by the SGP model were forced to be $\sim$ 9 min by the strongly constraining priors on the hyperparameters. This way the GP was ensured to model the short-period part of the solar granulation and p-mode timescale spectrum.

In summary, planetary radii are systematically overestimated if a flat baseline is applied. The overestimation bias becomes more significant with decreasing brightness. The noise model in our light curves contains noise with characteristic timescales of a few minutes up to five hours. Baseline models, which consider variations caused by noise sources at the shorter end of this range to be stochastic and to only account for variations with typical timescales of the order of hours, successfully mitigate this overestimation bias. However, the SHO-GP baseline, which was forced to model variations at the order of a a few tens of minutes, results in a similar bias as the flat baseline model. Taking into account all types of baseline models, no obvious correlation of the precision of the retrieved radius with the impact parameter can be observed. This indicates that even in the case of high impact parameters, the shorter transit duration does not negatively affect the precision of the retrieved radius. We note that all of these conclusions were drawn when imposing strong priors on the limb-darkening coefficients. When the limb-darkening coefficients are not well constrained and left free in the analyses, \citet{2016MNRAS.457.3573E} and \citet{Sulis2020} have shown that the strong degeneracy with the impact parameter can lead to biased planetary parameters.


\section{Transit detection}
\label{sec_detection}

The final light curve products contain a transit every 24 hours. In order to evaluate the performance of transit detection algorithms we created light curves containing only two individual transits spaced exactly one year apart. We started by removing all transits except for the first and the 45th transit by dividing the corresponding flux values by the transit shape. We then split the light curve exactly in the middle of the dataset and shifted the second half in such a way that the transit-mid-point of the 45th transit is placed exactly one year after the transit-mid-point of the first transit. Both parts of the light curve are then normalised by the mean of their first 300 data points. We then repeat this procedure for the next 43 transits in the original final light curve products, keeping every time only the n-th and the (n+44)-th transit. This results in 44 different light curves containing each a total of 88 days of data and two transits exactly one year apart. Our dataset therefore mimics a subset of the dataset that will actually be provided by PLATO, which will be at least two years of consecutive data. The subset provides a realistic dataset to test the performance of transit search algorithms, because it contains two distinct transit events and 87 days of out-of-transit (i.e. noise dominated) data and different transit events injected at different phases of the 3-month PLATO instrumental noise cycle. By doing this for all combinations of magnitudes and impact parameters, we end up with a total of 1100 light curves. 


Prior to performing a transit search, we corrected the light curves for instrumental systematics just as we would do for real observations. We performed sigma clipping and removed all points with the median absolute deviation (MAD) higher than 5 to discard outliers. Additionally, we fit a second order polynomial to each of the two sets of 44 consecutive days to remove the slope in the normalised flux during a single PLATO cycle. An example of an input light curve, the fitted slope, and a flattened light curve were published online \footnote{\url{https://zenodo.org/records/13938622}}. We developed our own transit search algorithm to sample the datasets for transit-like features. The algorithm mainly follows the approach of the transit least squares \texttt{Python} package \texttt{TLS}\footnote{\url{https://github.com/hippke/tls}}   \citep{Hippke2019} by applying a grid of transit models (see Table \ref{table_detection_parameters}) to the light curve and then comparing the resulting residuals with the residuals of the null-hypothesis (i.e. no transit). In order to make the algorithm more computationally efficient it only applied the transit model to sections of the light curve, where the mean of all data points within the smallest model transit-duration in the grid was more than $40$ ppm smaller than the overall mean of the dateset. We note that the algorithm was not provided any a priori knowledge on the transit location. The most important difference to the \texttt{transit least squares} algorithm is that transit durations are not evaluated in samples, but in time, which allows to also search light curves with gaps, varying cadences, and unknown numbers of transit events. We employed again a quadratic limb-darkening law using the same coefficients as in Sect. \ref{sec_detrending}. The algorithm was set-up to flag all events with a signal-to-noise ratio \citep[S/N, as defined by Equation 3 in][]{Kipping2023} of more than three and with two distinct transit events. A flagged event is rated as a correct detection if the transit-mid-times of both injected transit events are recovered within a quarter of a day of the correct values. If at least one of two transit-mid-times is retrieved correctly, the event is rated as a 1-transit-detection. In the case that a 1-transit-detection is not also a correct detection, the detection algorithm has correctly identified at least one of the transits but has paired the signal of this transit with a false positive signal somewhere else in the light curve (i.e. the transit-mid-time was recovered correctly but the period was wrong). In some of these cases the detection algorithm correctly recovered both transits but paired both of them with a different false positive signal, resulting in two flagged events, which both would be considered a 1-transit-detection. In our study these cases are counted as only one 1-transit-detection. All other flagged events are considered false positives. Because the out-of-transit data is identical in all the light curves that only vary the time of the injected transit except for the exclusion of the in-transit-data, false positive signals are usually flagged in almost all of these light curves. Such signals were only counted once, meaning that the total number of false positives represents the number of unique false positive signals per combination of impact parameter and stellar magnitude. The depth of the retrieved transit is not assessed.

\begin{table}
    \caption{Grid of transit model parameters used in the transit search algorithm.}
    \centering
    \begin{tabular}{cccc}
    \toprule
    \toprule
    Parameter & Limits & Step & Unit  \\
    \midrule
    Orbital period $P$ & $322 - 409$ & $0.2$ & days\\
    Transit mid-time $T_0$ & $0.0 - 44.0$ & $ 0.021$ & days\\
    Planetary radius $R_p$ & $0.9-2.0$&$0.1$ & $R_{\oplus}$ \\
    Impact parameter $b$ & $0.0-0.9$ & $0.1$ & - \\
    \bottomrule
    \end{tabular}
    \label{table_detection_parameters}
\end{table}

\begin{table*}
    \caption{Model parameters used in the transit analysis when trying to recover the planetary signal.}
    \centering
    \begin{tabular}{lcccc}
    \toprule
    \toprule
    Parameter & Symbol & Prior & Correct value \tablefootmark{(a)} &  Units \\
    \midrule
    \textbf{Planet Parameters}&&&& \\
    \quad Orbital period & $P$ & $\mathcal{U}(P_g - 0.5,P_g+0.5)$ & $365.25$ & days \\
    \noalign{\smallskip}
    \quad Transit time & $T_{0}$ & $\mathcal{U}(T_{0,g} - 0.5, T_{0,g} + 0.5)$ & 0.4995 - 0.5005 \tablefootmark{(b)} & days\\
    \quad Planet-to-star radius ratio & $R_{P}/R_\star$ & $\mathcal{U}(0.0,0.02)$ & $0.009168$  &- \\
    \quad Scale semi-major axis & $(R_\star + R_p) / a$ & $\mathcal{U}(0.0,0.02)$ &  $0.004693$ &  - \\
    \noalign{\smallskip}
    \quad Cosine of inclination & $\rm cos(i)$ & $\mathcal{U}(0.0,0.006)$ &  varying &  - \\
    \quad White noise term & $\sigma$ & $\mathcal{U}(45,6738)$ &  varying &  ppm \\
    \noalign{\smallskip}
    \midrule
    \textbf{Stellar Parameters} &&&& \\
    \quad Stellar mass & $M_\star$ &$\mathcal{N}(1.00,0.05)$ & 1.0 &  $M_\odot$ \\
    \quad Stellar radius & $R_\star$ &$\mathcal{N}(1.000,0.015)$ & 1.0 &  $R_\odot$ \\
    \multirow{2}{*}{\quad limb-darkening coefficients} & $q_1$ &$\mathcal{N}(0.392,0.05)$ & - &  - \\
    \noalign{\smallskip}
    & $q_2$ & $\mathcal{N}(0.322,0.05)$ & - &  - \\
    \midrule
    \textbf{SHO GP} &&&& \\
    \quad Scaled amplitude & $\mathrm{ln(S_0)}$ &$\mathcal{N}(-26.79,0.04)$ & - & - \\
    \quad Quality factor & $\mathrm{ln(Q)}$ &$\mathcal{N}(-0.402,0.025)$ & - & - \\
    \quad Angular Frequency & $\mathrm{ln}(\omega_0)$ &$\mathcal{N}(6.932,0.011)$ & - & - \\
    \bottomrule
    \end{tabular}
    \tablefoot{Initial guess values retrieved by the previous transit search algorithm a denoted by a lower case g. The Gaussian priors with mean $\mu$ and standard deviation $\sigma$ are displayed as $\mathcal{N}(\mu, \sigma)$. $\mathcal{U}(U_l,U_u)$ represents an uniform prior with $U_l$ and $U_u$ as lower and upper limits respectively.\\
    \tablefoottext{a}{Value of the injected transit signal.}\\
    \tablefoottext{b}{The injected transit mid-time is always the exact middle of the dataset, which varies across the different observations according to the exact observation times of the SDO images.}}
    \label{tab_fitting_parameters}
\end{table*}

\begin{figure}
    \centering
    \includegraphics[width=\hsize]{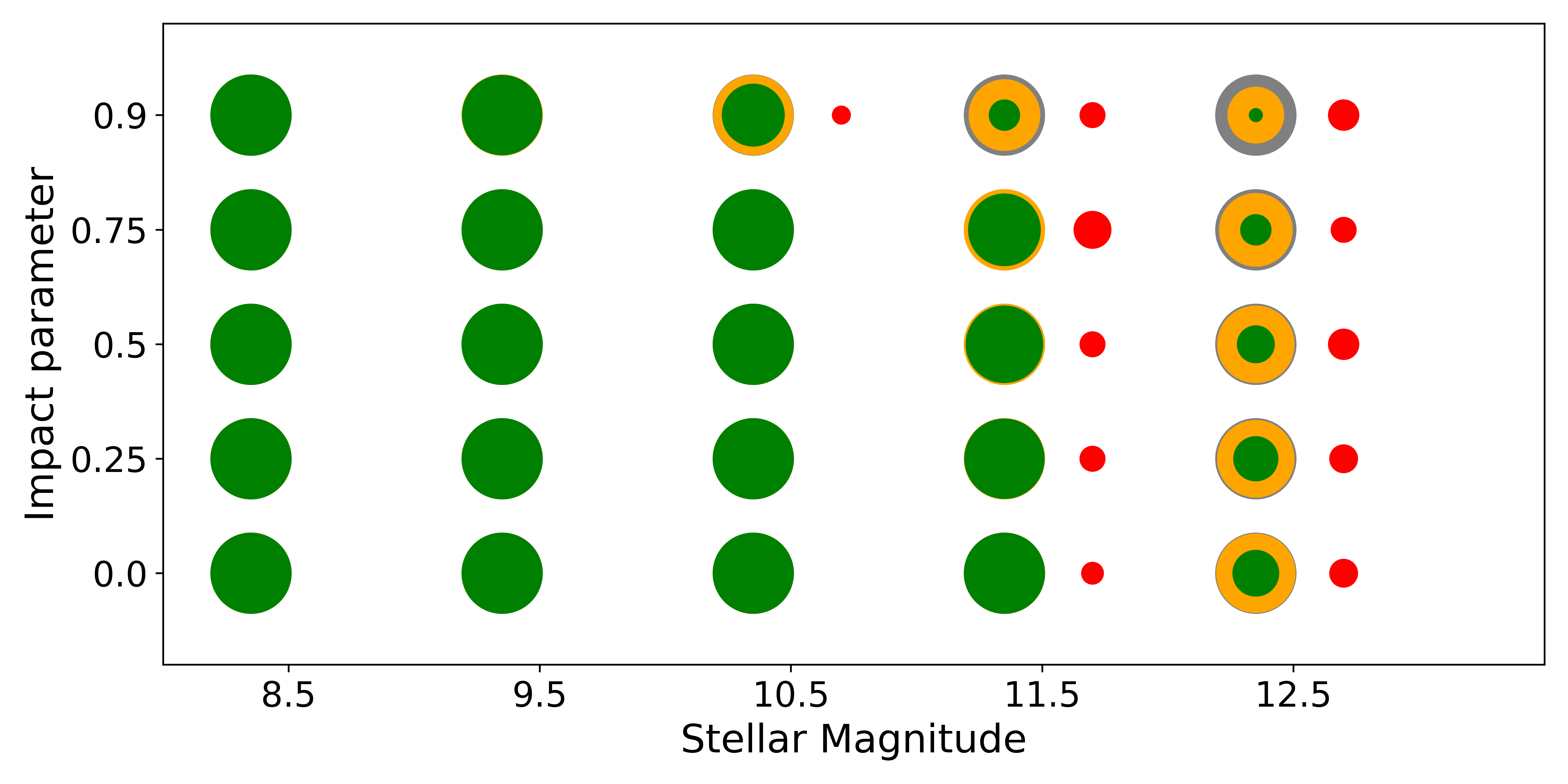}
    \caption{Overview of detections and false positives of the transit search algorithm as a function of the stellar magnitude and impact parameter. Grey circles mark the number of injected transit events, green circles the number of correct detections, orange circles the number of events where at least one of the two injected transits was detected, and red circles the number of unique false positives. Numbers are proportional to the area of a circle.}
    \label{fig_detection}
\end{figure}

Fig. \ref{fig_detection} illustrates the results of the transit search algorithm as a function of magnitude and impact parameter. Grey circles mark the number of injected transit events, green circles the number of correct detections, orange circles the number of 1-transit-detections, and red circles the number of unique false positive signals. In 1006 (91.5\%) out of the 1100 injected cases at least one of the transits was correctly recovered. 852 (77.5\%) cases are rated as a correct detection, meaning that both the transit-mid-time and the period have been correctly identified. A total of 52 unique false positives were also flagged. For the brightest target (8.5 mag) all injected transits are correctly recovered. In the case of the 9.5 mag target, all except for one event are rated as correct detections. For the one remaining event only one of the two injected transits has been successfully recovered. However, it must be noted that at the specific position of the missed second transit, the HMI data has a 6-hour data gap which in the case of the missed transit at impact parameter $b=0.9$ covers the majority of the transit event leaving only roughly 30 minutes of the ingress. Therefore the missed detection of this transit is not due to a failure of the detection algorithm but due to a lack of data. Therefore we conclude that also for 9.5 mag targets all of the injected transit events were correctly recovered. For the 10.5 mag targets all events with impact parameters $b<0.9$ were rated as correct detections. Out of the 44 events with impact parameters $b=0.9$, 26 were also marked as correct detections, while of the remaining 18 events, 17 were rated as a 1-transit-detection. The one completely undetected event again corresponds to the one event where in the case of impact parameter $b=0.9$, due to a lack of HMI data, only roughly 30 minutes of the ingress of the second transit are available. Therefore we conclude that if for a 10.5 mag target at least two distinct transit events are present in the data, at least one of them is always detected. In the case of the 10.5 mag target also two false positive signals were flagged. However, if at least one transit was correctly detected in a light curve, the S/N of this correct signal was always higher than the S/N of the false positive signal. For faint targets (11.5 and 12.5 mag), the detection rate decreases with increasing impact parameter. However, for the majority of the injected events at least one of the transits was successfully recovered with a 1-transit-detection rate of 95.5\% and 81.4\% respectively. We also observed several false positive signals being flagged in the case of faint targets.

\begin{table*}
    \caption{Overview of the PLATO-like transit fitting results.}
    \centering
    \begin{tabular}{cc|cccccc}
    \toprule
    \toprule
     & Magnitude & Correct & 1-Transit  & False &Discarded&  Median $R_{\rm pl}$ [$R_{\oplus}$] & Median $\Delta R_{\rm pl}$ [$R_{\oplus}$]\\
    \midrule
    \multirow{4}{5mm}{SP}&$8.5$&$215$&$0$&$0$&$0$&$0.99\pm0.02$&$0.02$ \\ 
&$9.5$&$215$&$0$&$0$&$0$&$0.99\pm0.03$&$0.03$ \\ 
&$10.5$&$198$&$17$&$0$&$0$&$1.00\pm0.05$&$0.05$ \\ &$11.5$&$160$&$30$&$0$&$25$&$1.01\pm0.08$&$0.09$ \\  
    \midrule 
    \multirow{4}{5mm}{SGP}&$8.5$&$215$&$0$&$0$&$0$&$0.98\pm0.02$&$0.02$ \\ 
&$9.5$&$215$&$0$&$0$&$0$&$0.99\pm0.04$&$0.03$ \\ 
&$10.5$&$198$&$16$&$0$&$1$&$0.99\pm0.05$&$0.05$ \\ 
&$11.5$&$161$&$30$&$0$&$24$&$1.01\pm0.09$&$0.09$ \\ 
    \bottomrule
    \end{tabular}
    \tablefoot{Per magnitude and noise model a set of 215 transit pairs was fitted. Each set contains transit pairs with impact parameters varying between 0.0 and 0.9. Detections that recover both injected transit-mid times within 0.25 days of the correct values are considered correct. Detections that only recover one of the two injected transit-mid times within 0.25 days are  flagged as 1-Transit detections. Detections for which the planetary radius is measured with less than 3$\sigma$-confidence are discarded. All others are considered false positive detections. The medians of the derived planetary radius $R_{\rm pl}$ and its uncertainty $\Delta R_{\rm pl}$ are calculated using only the correct solutions. The median of $R_{\rm pl}$ is stated including its 1$\sigma$-confidence interval. }
    \label{table_fitting_results}
\end{table*}

\section{PLATO-like transit fitting}
\label{sec_fitting}
We finally aim at testing the potential of PLATO to detect and characterise an Earth analogue by combining the information gained in Sect. \ref{sec_detrending} and \ref{sec_detection}. To this end we use the same light curves presented in Sect. \ref{sec_detection}, containing a total of 88 days of data and two individual transit events placed exactly one year apart. We first run the transit search algorithm introduced in Sect. \ref{sec_detection} and then, in the case of a flagged detection, fit for the solution with the highest signal-to-noise ratio. When fitting we use wide uniform priors on all parameters of the transit model centred around the value retrieved by the transit search algorithm. In order to impose realistic priors on the stellar parameters, we inflate the widths of the priors on the limb-darkening coefficients and add Gaussian priors on stellar radius and mass. For the limb-darkening coefficients, we assume uncertainties of $0.05$, which is a reasonable estimate considering the abilities of recent python routines such as \texttt{LDCU}\footnote{\texttt{LDCU} (\url{https://github.com/delinea/LDCU}) is a modified version of the python routine implemented by \citet{espinoza2015} that computes the limb-darkening coefficients and their corresponding uncertainties using a set of stellar intensity profiles accounting for the uncertainties on the stellar parameters. The stellar intensity profiles are generated based on two libraries of synthetic stellar spectra: ATLAS \citep{Kurucz1979} and PHOENIX \citep{husser2013}} to compute the limb-darkening coefficients and their uncertainties from theoretical stellar models. \citet{Bonfanti2023} for example present $u_1 = 0.300 \pm 0.010$ and $u_2 = 0.275 \pm 0.006$ for observations of the G-type star TOI-1055 (V-mag 8.7) in the TESS bandpass and \citet{Krenn2024} $q_1 = 0.49 \pm 0.05$ and $q_2 = 0.359 \pm 0.015$ for the G-type star TOI-421 (V-mag 9.9). With respect to stellar radius and mass, the PLATO mission is dedicated to provide precise stellar parameters supporting its exoplanet science. It will do this by both using its astroseismology programme and operating its own extensive ground-based follow-up campaign. The goal is to provide stellar radii at a precision of 2\% and stellar masses with an accuracy of 10\% at least \citep{Rauer2024}. We adopt uncertainties on stellar radius and mass of 1.5\% and 5\% respectively, which are consistent with the PLATO science goal and also have already previously been achieved even for faint G-type stars. For example \citet{Tregloan-Reed2013} report $R_{\star} = 1.004 \pm 0.016 \, R_{\odot}$ and $M_{\star} = 0.90 \pm 0.04 \, M_{\odot}$ for the G-type star WASP-19 (V-mag 12.3).

We only include data points that are within 0.5 days of the expected transit-mid-point in the transit analysis. Because the PLATO core sample is comprised of bright targets (V-mag < 11) we did not consider light curves from the target star with stellar magnitude 12.5. We include all five impact parameters $b = 0, 0.25, 0.5, 0.75$, and $0.9$. We discard the light curves which include a transit on the 26th of December 2014, because due to a gap in the HMI data it misses six hours of in-transit data. In this way we end up with a total of 860 unique pairs of transits. We run every transit fit twice, once employing a spline baseline and once employing a SHO-GP. These two baseline models were chosen, because they model variations with different characteristic timescales. While the SP baseline model only models variations with characteristic timescales of a few hours, SGP is ensured to model variations with characteristic timescale of a few minutes by strongly constraining the GP hyperparameters analogous to the analysis performed in Sec. \ref{sec_detrending}. In this way we end up with a total of 1720 individual transit fits. An overview of all model parameters and imposed priors is given in Table \ref{tab_fitting_parameters}. An example of a fitted transit light curve for a 9.5 mag star using both the SP and the SGP baseline models, as well as assuming a flat baseline, was made available online \footnote{\url{https://zenodo.org/records/13938622}}.

Table \ref{table_fitting_results} presents the results of realistic PLATO-like transit search and fitting. Solutions for which the planetary radius is measured with less than 3$\sigma$-confidence are discarded. 
Analogous to Sect. \ref{sec_detection} an event is rated as a correct detection if the transit-mid-times of both injected transit events are recovered within a quarter of a day of the correct values. If at least one of two transit-mid-times is retrieved correctly, the event again is rated as a 1-transit-detection. All other solutions are rated as false positives. 

We do find that all variations of the algorithm perform extraordinary well. For bright targets 100\% of the injected transits are recovered correctly. In the case of the magnitude 10.5 target, there are a few instances in which the transit search algorithm paired a single correct transit with an incorrect second signal. This results in the recovery of the correct transit-mid-time, but a wrong period. For the faintest target (magnitude 11.5) the majority of transits are still recovered correctly. Again we do see some single transit detections, but this time also a few solutions have been discarded because the required $3\sigma$ radius precision was not reached. Not a single false positive detection was flagged. This leads us to the conclusion that the combination of a transit search algorithm with a transit-fitting algorithm can ensure reliable detections even for faint targets. With respect to the accuracy of the retrieved planetary radii of correct detections, we see that independently of the magnitude the median of the recovered radii is always within 1$\sigma$ of the injected value. We find that the 3\% radius precision goal set by the PLATO mission for Earth-like planets orbiting bright stars is achieved for both the 8.5 and the 9.5 mag targets. We do not find any statistically significant difference between the SP and the SGP baseline models. While the SP model only accounts for variability with typical timescales of a few hours, SGP models variability at the order of $\sim 9$ min. The fact that they perform equally well at retrieving the correct planet radius indicates that the SHO-GPs treatment of shorter term variability does not improve the precision of the radius measurement. On the other hand, there is no more overestimation bias for the retrieved radii of the SGP model, as it was observed when only fitting a single transit event in Sec. \ref{sec_detrending}. This implies that the combination of two transit events is sufficient to adequately resolve the correlation between planet size, impact parameter, and stellar variability.


\section{Conclusions}
We aimed to quantify the impact of short-term stellar variability on PLATO's prime science case - the detection and characterisation of Earth analogues - by using the Sun as a proxy. To this extend we simulated PLATO-like light curves containing the transit of an Earth analogue by injecting Earth-like transits in solar data of SDO's HMI instrument and adding a noise model employing the \PlatoSim simulator assuming an observation by all 24 normal cameras. To remove the flux variations caused by the orbital motion of the recording satellite we removed any variability with characteristic timescales longer than five hours - including variability caused by magnetic features on the solar surface - from the HMI data by applying a smooth Savitzky-Golay filter \citep{Savitzky1964} before adding the transit and noise models. We used 88 consecutive days of HMI observations and simulated PLATO observations for five different impact parameters and five different stellar magnitudes. When trying to recover the injected transit signals, we find that planetary radii are systematically overestimated if a flat baseline is applied. Baseline models, which consider variations caused by noise sources with characteristic timescales of a few minutes to be stochastic and only account for variations with typical timescales of the order of hours, successfully mitigate this overestimation bias. However, the SHO-GP baseline, which was made to model variations at the order of a few minutes, results in similar overestimation as the flat baseline model. We also find that if the limb-darkening coefficients of the host star are properly constrained, a high impact parameter and therefore a shorter transit duration, does not negatively affect the precision of the retrieved planetary radius.

When employing a transit search algorithm similar to the well-known transit least squares algorithm, we find that transit signals can reliably be detected with a high signal-to-noise ratio for bright targets (8.5, 9.5, and 10.5 mag). For faint targets (11.5 and 12.5 mag) a correct detection is highly likely, but the possibility of a false positive detection can not be excluded using the tools presented in this study. Again the impact parameter does not negatively affect the ability to detect the transit signal except for the highest value in our sample ($b = 0.9$). When also fitting for the transit parameters using realistic priors on stellar and planetary parameters, we find that both simple (e.g. spline) and more complex models (e.g. using GPs) are suitable to mitigate the effects of stellar and instrumental noise. 

For magnitudes between 8.5 and 10.5 mag, which corresponds to the PLATO core sample, solar-like short-term variability will not affect PLATO's capability to detect a transit signal of an Earth analogue and accurately determine the transit shape at high-precision as long as it has observed at least two distinct transit events and strong priors are used to constrain the stellar limb-darkening coeffiencts. Even for fainter targets (11.5 mag) a detection and correct sizing is possible, but transit search algorithms have to be combined with transit-fitting algorithms to avoid false positive detections being flagged. With regard to the goal set by the PLATO mission to measure precise and accurate radii of Earth-like planets orbiting bright stars, we find that this is a realistic goal for stars with magnitudes $\leq$ 9.5 if the characteristics of the host star, including its activity-induced photometric variability, are well understood. Precise stellar radii and limb-darkening coefficients are necessary to precisely determine the planets size. Additional information constraining the variability due to oscillations and granulation of solar-like stars will also aid the precise measurement of planetary radii by imposing additional priors on models trying to mitigate the effects of flux variations caused by stellar variability. 

\section*{Data availability}
The set of artificial PLATO-like light curves was made available online at \url{https://zenodo.org/records/13939257}. Supplementary figures were made available online at \url{https://zenodo.org/records/13938622}.

\begin{acknowledgements}
This work presents results from the European Space Agency
(ESA) space mission PLATO. The PLATO payload, the PLATO Ground
Segment and PLATO data processing are joint developments of ESA and
the PLATO mission consortium (PMC). Funding for the PMC is provided at national levels, in particular by countries participating in the PLATO Multilateral Agreement (Austria, Belgium, Czech Republic, Denmark, France, Germany, Italy, Netherlands, Portugal, Spain, Sweden, Switzerland, Norway, and United Kingdom) and institutions from Brazil. Members of the PLATO Consortium can be found at https://platomission.com/. The ESA PLATO mission website is https://www.cosmos.esa.int/plato. We thank the teams working for PLATO for all their work.
This work has been carried out within the framework of the NCCR PlanetS supported by the Swiss National Science Foundation under grants 51NF40\_182901 and 51NF40\_205606. 
ML acknowledges support of the Swiss National Science Foundation under grant number PCEFP2\_194576.
SS acknowledges support from the ``Programme National de Physique Stellaire`` (PNPS) and ``Programme National de Planétologie`` (PNP) of CNRS/INSU co-funded by CEA and CNES. MD and SS gratefully acknowledge financial support from CNES, focused on the PLATO mission. 
NJ, JDR, and DS thank the Belgian Federal Science Policy Office (BELSPO) for the provision of financial support in the framework of the PRODEX Programme of the European Space Agency (ESA) under contract number PEA 4000137604.
\end{acknowledgements}

\bibliographystyle{aa} 
\bibliography{bibliography} 

\begin{appendix} 
\onecolumn
\section{Mathematical description of used GP models}
\label{sec_appendix_gp}

The Matérn-3/2 GP kernel is defined by a length scale parameter $\rho$ \citep{Rasmussen2006}:

\begin{equation}
    k_{\text{Matérn-3/2}}(\tau,\rho) = \left(1 + \frac{\sqrt{3} \tau}{\rho}\right) e^{-\frac{\sqrt{3} \tau}{\rho}}.
\end{equation}

The SHO-GP kernel is defined by an amplitude of variations $S_0$, an angular frequency of the variations $\omega_0$, and a quality factor of the oscillator $Q$ \citep{Foreman-Mackey2017}:

\begin{equation}
k_{\text{SHO}}(\tau; S_0, Q, \omega_0) = S_0 \, \omega_0 \, Q \, e^{-\frac{\omega_0 \, \tau}{2Q}}
\begin{cases}
  \cosh(\eta \, \omega_0 \, \tau) + \frac{1}{2\eta \, Q} \sinh(\eta \, \omega_0 \, \tau), & 0 < Q < \frac{1}{2} \\
  2(1 + \omega_0 \, \tau)&Q = \frac{1}{2} \\
  \cos(\eta \, \omega_0 \, \tau) + \frac{1}{2\eta \, Q} \sin(\eta \, \omega_0 \, \tau), & \frac{1}{2} < Q
\end{cases}
\end{equation}

\noindent with $\eta = \left| 1 - \left(4Q^2\right)^{-1} \right|^{1/2}$.

\section{Results of the comparison of different baseline models}

\begin{table}[!ht]
    \caption{Overview of the distributions of retrieved planetary radii when comparing different baseline models.}
    \centering
    \begin{tabular}{cc|ccccc}
    \toprule
    \toprule
    Magnitude & Impact Parameter & FB & P2 &  SP & MGP & SGP \\
    \midrule
    \multirow{6}{5mm}{8.5} &0.0&$1.08\pm0.07$ &$1.00\pm0.07$ &$1.00\pm0.08$ &$1.04\pm0.06$ &$1.08\pm0.07$ \\ 
    &0.25&$1.08\pm0.07$ &$1.02\pm0.06$ &$1.02\pm0.06$ &$1.04\pm0.06$ &$1.07\pm0.07$ \\ 
    &0.5&$1.06\pm0.07$ &$1.01\pm0.07$ &$1.00\pm0.07$ &$1.04\pm0.06$ &$1.05\pm0.07$ \\ 
    &0.75&$1.08\pm0.08$ &$0.99\pm0.08$ &$0.99\pm0.08$ &$1.04\pm0.08$ &$1.08\pm0.08$ \\ 
    &0.9&$1.07\pm0.14$ &$1.01\pm0.07$ &$1.02\pm0.07$ &$1.02\pm0.09$ &$1.06\pm0.13$ \\ 
    &\textbf{all}&$\mathbf{1.07\pm0.07}$ &$\mathbf{1.01\pm0.07}$ &$\mathbf{1.01\pm0.07}$ &$\mathbf{1.04\pm0.06}$ &$\mathbf{1.07\pm0.07}$ \\ 
    \midrule 
    \multirow{6}{5mm}{9.5}&0.0&$1.17\pm0.13$ &$0.99\pm0.12$ &$0.99\pm0.13$ &$1.03\pm0.08$ &$1.17\pm0.13$ \\ 
    &0.25&$1.14\pm0.14$ &$1.03\pm0.10$ &$1.04\pm0.10$ &$1.03\pm0.07$ &$1.15\pm0.13$ \\ 
    &0.5&$1.11\pm0.12$ &$0.99\pm0.09$ &$1.00\pm0.10$ &$1.00\pm0.08$ &$1.12\pm0.12$ \\ 
    &0.75&$1.11\pm0.12$ &$1.03\pm0.08$ &$1.03\pm0.07$ &$1.02\pm0.08$ &$1.11\pm0.11$ \\ 
    &0.9&$1.03\pm0.19$ &$1.03\pm0.12$ &$1.02\pm0.11$ &$1.01\pm0.10$ &$1.02\pm0.15$ \\ 
    &\textbf{all}&$\mathbf{1.14\pm0.13}$ &$\mathbf{1.02\pm0.11}$ &$\mathbf{1.02\pm0.10}$ &$\mathbf{1.02\pm0.08}$ &$\mathbf{1.13\pm0.12}$ \\ 
    \midrule 
    \multirow{6}{5mm}{10.5}&0.0&$1.26\pm0.22$ &$1.05\pm0.18$ &$1.04\pm0.20$ &$1.04\pm0.09$ &$1.26\pm0.20$ \\ 
    &0.25&$1.31\pm0.24$ &$1.07\pm0.18$ &$1.08\pm0.20$ &$1.05\pm0.11$ &$1.31\pm0.24$ \\ 
    &0.5&$1.24\pm0.21$ &$1.06\pm0.20$ &$1.07\pm0.19$ &$1.01\pm0.10$ &$1.24\pm0.18$ \\ 
    &0.75&$1.27\pm0.20$ &$1.04\pm0.17$ &$1.05\pm0.14$ &$1.01\pm0.13$ &$1.26\pm0.24$ \\ 
    &0.9&$1.18\pm0.25$ &$1.10\pm0.24$ &$1.09\pm0.21$ &$0.96\pm0.21$ &$1.18\pm0.24$ \\ 
    &\textbf{all}&$\mathbf{1.25\pm0.20}$ &$\mathbf{1.06\pm0.17}$ &$\mathbf{1.06\pm0.18}$ &$\mathbf{1.03\pm0.10}$ &$\mathbf{1.26\pm0.20}$ \\ 
    \midrule 
    \multirow{6}{5mm}{11.5}&0.0&$1.46\pm0.36$ &$1.05\pm0.25$ &$1.04\pm0.24$ &$1.12\pm0.17$ &$1.44\pm0.40$ \\ 
    &0.25&$1.55\pm0.47$ &$0.97\pm0.32$ &$0.99\pm0.25$ &$1.01\pm0.14$ &$1.54\pm0.45$ \\ 
    &0.5&$1.55\pm0.47$ &$1.15\pm0.34$ &$1.11\pm0.31$ &$1.05\pm0.14$ &$1.50\pm0.39$ \\ 
    &0.75&$1.42\pm0.28$ &$1.10\pm0.25$ &$1.10\pm0.23$ &$1.05\pm0.19$ &$1.43\pm0.27$ \\ 
    &0.9&$1.33\pm0.37$ &$1.02\pm0.21$ &$0.98\pm0.26$ &$0.93\pm0.14$ &$1.36\pm0.35$ \\ 
    &\textbf{all}&$\mathbf{1.46\pm0.34}$ &$\mathbf{1.06\pm0.26}$ &$\mathbf{1.05\pm0.26}$ &$\mathbf{1.05\pm0.16}$ &$\mathbf{1.47\pm0.34}$ \\ 
    \midrule 
    \multirow{6}{5mm}{12.5}&0.0&Upp. boundary &$0.97\pm0.41$ &$1.01\pm0.46$ &$1.28\pm0.30$ &$1.66\pm0.41$ \\ 
    &0.25&Upp. boundary &$0.96\pm0.60$ &$0.97\pm0.60$ &$1.25\pm0.20$ &$1.61\pm0.35$ \\ 
    &0.5&Upp. boundary &$0.96\pm0.47$ &$0.98\pm0.45$ &$1.18\pm0.38$ &$1.82\pm0.45$ \\ 
    &0.75&Upp. boundary &$1.05\pm0.69$ &$1.05\pm0.71$ &$1.12\pm0.29$ &$1.60\pm0.33$ \\ 
    &0.9&Upp. boundary &$0.90\pm0.56$ &$0.91\pm0.52$ &$1.05\pm0.37$ &$1.72\pm0.42$ \\ 
    &\textbf{all}&$\textbf{Upp. boundary}$ &$\mathbf{0.98\pm0.53}$ &$\mathbf{0.99\pm0.54}$ &$\mathbf{1.21\pm0.31}$ &$\mathbf{1.67\pm0.35}$ \\ 
    \bottomrule
    \end{tabular}
    \tablefoot{The medians and standard deviations of 88 retrieved planetary radii $R_{pl}$ for different combinations of stellar magnitudes, impact parameters, and baseline models are shown. Upp. Boundary indicates that the majority of fitting runs ended up at or near the upper boundary of the uniform prior constraining the parameter.}
    \label{table_detrending_results}
\end{table}

\end{appendix}
\end{document}